\documentstyle[12pt]{article}
\def\be{\begin{equation}}
\def\ee{\end{equation}}
\def\ba{\begin{eqnarray}}
\def\ea{\end{eqnarray}}

\def\fun#1#2{\lower3.6pt\vbox{\baselineskip0pt\lineskip.9pt

\ialign{$\mathsurround=0pt#1\hfill##\hfil$\crcr#2\crcr\sim\crcr}}}
\begin{document}

\setbox16= \hbox{der \kern -17.2pt \raise %
 3.8pt \hbox{-} \vrule width -2.9pt depth 0pt }
\def\dj{\copy16}

\begin{titlepage}
\null\vspace{-62pt}
\begin{flushright}LBL-35530;\ UCB-PTH-94/09\\
      April 1994

\end{flushright}
\vspace{0.1in}
\centerline{{\large \bf Approximate Flavor Symmetries }$^{1,2}$}
\vspace{0.2in}
\centerline{
Andrija Ra\v{s}in $^{3}$}
\vspace{0.2in}
\centerline{{\it Department of Physics,
University of California, Berkeley, CA 94720}}
\centerline{{\it and Theoretical Physics Group, Lawrence Berkeley Laboratory}}
\centerline{{\it University of California, Berkeley, CA 94720}}
\vspace{.5in}
\baselineskip=22pt

\centerline{\bf Abstract}
\begin{quotation}
We discuss the idea of approximate flavor
symmetries. Relations between approximate flavor symmetries
and natural flavor conservation and democracy models
is explored. Implications for neutrino
physics are also discussed.

\vspace{2.5in}

\baselineskip=15pt

----------------------------------------------------------------------------------------------

$^{1}$ Short talk given at the Second IFT workshop on Yukawa couplings
and the origin of mass, Feb 11-13, University of Florida, Gainesville.

$^{2}$ This work was supported in part by the Director, Office of
Energy Research, Office of High Energy and Nuclear Physics, Division of
High Energy Physics of the U.S. Department of Energy under Contract
DE-AC03-76SF00098.

$^{3}$ On leave of absence
from the Ru\dj \hspace{.1in} Bo\v{s}kovi\'c Institute, Zagreb, Croatia.
Address after September 1, 1994: Department of Physics,
University of Maryland, College Park 20742.
\end{quotation}
\end{titlepage}

\newpage
\setcounter{page}{1}

\baselineskip=24pt

\section{Introduction}
\label{intro}

In the Standard Model, Yukawa couplings $\lambda$ are
defined as couplings between fermions and Higgs scalars:

\begin{equation}
{\cal L}_Y =
         ( {\lambda}^{U^a}_{ij} {\bar Q}_i U_j \frac {\tilde{H}_a} {\sqrt{2}} +
            \lambda^{D^a}_{ij} {\bar Q}_i D_j \frac {H_a} {\sqrt{2}}+
           \lambda^{E^a}_{ij} {\bar L}_i E_j   \frac {H_a} {\sqrt{2}} + h.c.),
\label{yuk}
\end{equation}
where $Q_i$ and $L_i$ are $SU(2)$ doublet quarks and leptons while $U_i$ ,
$D_i$ and $E_i$ are $SU(2)$ singlets and $i = 1,2,3$ is a generation label.
$H_a$ are Higgs SU(2) doublets, and $a=1,\ldots,n$, where $n$ is the number of
Higgs doublets.

The Standard Model can easily be extended to
accommodate neutrino masses. We can
introduce higher dimensional operators $\frac {1} {M} H H L_i L_j$ which give
Majorana neutrino masses. We can also add SU(2) singlet neutrinos which can
also have Majorana masses or Dirac masses similar
to (\ref{yuk}). Neutrino mass is then in general
described by a $6 \times 6$ Yukawa matrix.

As opposed to, e.g. couplings of fermions and vector bosons, Yukawa
couplings in the Standard Model are not constrained
nor related to each other by any symmetry
principle; i.e., they enter the Lagrangian as arbitrary complex numbers.
These numbers are then only fixed by experiment, namely by measuring
fermion masses and mixing angles.

Several classes of models/ans\"{a}tze for Yukawa couplings exist today,
and we list them below:

\noindent $\bullet$ Approximate flavor
symmetries\cite{frog79}-\cite{leur93},
in which the entries in the Yukawa matrices are entered as small
parameters by which the flavor symmetries are broken.

\noindent $\bullet$ Fritzsch and/or GUT inspired
models\cite{frit77}-\cite{kapl94}, in which some entries in the Yukawa
mass matrices are assumed to be zero (e.g. by discrete flavor symmetry), and
others may be related by some GUT relation.

\noindent $\bullet$ Flavor democracy models\cite{hara77}-\cite{frit94},
in which all
the entries in the Yukawa matrices are equal (i.e., no flavor symmetry),
and hierarchy comes from diagonalization and RGE running.

\noindent $\bullet$ String inspired models, composite models, ...

This is in no way a complete list, of course. Also, most of the
work done actually falls into several categories above, showing that there are
common ideas, which will hopefully lead us  to a certain trail beyond the
Standard Model.

In this talk, we will concentrate on approximate flavor symmetries,
since they are relatively simple and model independent.

\section{Approximate flavor symmetries}

In the Standard Model, the gauge interactions of the fermions
\begin{equation}
{\cal L}_0 = i \bar {Q}\!\!\not\!\!D Q + i \bar {U}\!\!\not\!\!D U +
      i \bar {D}\!\!\not\!\!D D + i \bar {L}\!\!\not\!\!D L +
      i \bar {E}\!\!\not\!\!D E
\end{equation}
have global flavor symmetries and these are broken by Yukawa couplings
(\ref{yuk}). We can understand the Yukawa couplings as being
naturally small\cite{thoo80}, because in the limit that they become
zero, the theory gets a larger (flavor) symmetry. This is certainly
warranted by the smallness of fermion masses (except the top),
which arise from Yukawa couplings.

One of the simplest assumptions one can make is that each of the chiral
fermion fields carries a flavor symmetry which is broken by a small
parameter, which we will call $\epsilon$. For example, Froggatt and
Nielsen\cite{frog79} think of $\epsilon$ as $\epsilon \approx \frac
{< \Phi_1 >} {< \Phi_0 >}$, where $< \Phi_1 >$ is the vev of
the scalar that breaks the flavor symmetry, and $< \Phi_0 >$ is
the vev of a superheavy field, therefore making $\epsilon$ small.
Thus, for example

\begin{equation}
\lambda^Q_{ij} \approx \epsilon_{Q_i} \epsilon_{U_j}
\label{epsi}
\end{equation}
where $\epsilon_{Q_i}$ is the breaking parameter of the flavor carried
by $Q_i$, and $\epsilon_{U_j}$ is the breaking parameter of the flavor carried
by $U_j$. Froggatt and Nielsen\cite{frog79},
as well as Leurer, Nir and Seiberg\cite{leur93} use
one or two $\epsilon$s, which enter the Yukawa matrices with different
powers, thus explaining the hierarchy of masses. We rather keep different
$\epsilon$s for different flavors, as it keeps the discussion more
model independent. Notice that in (\ref{epsi}) we assumed that the Higgs
field does not carry any flavor symmetry. If it did, then the Yukawa
couplings would be multiplied by another $\epsilon$ for the broken symmetry
carried by the Higgs field.

What do we know about $\epsilon$s? Although their approximate value
can be fixed by the known masses and mixings (at least in the quark
sector\cite{hall93}; lepton sector is much more speculative as the
neutrino masses and mixings are not known\cite{rasi94}; we will talk more
on this later), their
exact values depend on the underlying
theory, which we do not know. Therefore, relation like (\ref{epsi})
is not meant
to be an exact relation but rather an order of magnitude estimate, and we
are interested here mainly in general features, rather than specific
predictions.
It was shown that flavor changing interactions, with couplings
determined by approximate flavor symmetries, can involve new
scalars at the scale as low as the weak scale, and still satisfy
stringent experimental limits(\cite{anta92},\cite{hall93}).
This is opposed to the common view
that, for example, $K_L - K_S$ mass difference, implies high bounds on
the scale of new interactions, typically $\sim 1000 \, {\rm Te\!V}$.
However, it is precisely because of the approximate flavor symmetries
that the couplings of the new scalars are small,
lowering the naive bound considerably.

\section{Many Higgs doublet model}
\label{fsqse}

In this section we discuss the case of the minimal Standard
Model extended only by the addition of an arbitrary number of Higgs doublets.
In this case it is already known that, for the special case of Fritzsch-like
Yukawa matrices, the additional scalars need not be heavier than a TeV
\cite{chen87}. However, our results are independent of the particular
texture and depend only on the approximate flavor symmetry.

Let us look at a two Higgs doublet model (the generalization to many
Higgs doublets is trivial). For example, the up quark Yukawa couplings are

$\left( \begin{array}{ccc}
\bar{Q_1} & \bar{Q_2} & \bar{Q_3}
\end{array} \right)
\left( \begin{array}{ccc}
\sim \epsilon_{Q_1} \epsilon_{U_1} & \sim \epsilon_{Q_1} \epsilon_{U_2}
& \sim \epsilon_{Q_1} \epsilon_{U_3}  \\
\sim \epsilon_{Q_2} \epsilon_{U_1} & \sim \epsilon_{Q_2} \epsilon_{U_2}
& \sim \epsilon_{Q_2} \epsilon_{U_3}  \\
\sim \epsilon_{Q_3} \epsilon_{U_1} & \sim \epsilon_{Q_3} \epsilon_{U_2}
& \sim \epsilon_{Q_3} \epsilon_{U_3}
\end{array} \right)
\left( \begin{array}{c}
U_1 \\
U_2 \\
U_3
\end{array} \right) \frac {\tilde{H_1}} {\sqrt {2}}$

$+ \;\;  \left( \begin{array}{ccc}
\bar{Q_1} & \bar{Q_2} & \bar{Q_3}
\end{array} \right)
\left( \begin{array}{ccc}
\sim \epsilon_{Q_1} \epsilon_{U_1} & \sim \epsilon_{Q_1} \epsilon_{U_2}
& \sim \epsilon_{Q_1} \epsilon_{U_3}  \\
\sim \epsilon_{Q_2} \epsilon_{U_1} & \sim \epsilon_{Q_2} \epsilon_{U_2}
& \sim \epsilon_{Q_2} \epsilon_{U_3}  \\
\sim \epsilon_{Q_3} \epsilon_{U_1} & \sim \epsilon_{Q_3} \epsilon_{U_2}
& \sim \epsilon_{Q_3} \epsilon_{U_3}
\end{array} \right)
\left( \begin{array}{c}
U_1 \\
U_2 \\
U_3
\end{array} \right) \frac {\tilde{H_2}} {\sqrt {2}}$

\noindent and similarly for down type quark matrices, keeping in mind that each
entry in the matrices is uncertain by a factor of 2 or 3 (denoted by $\sim$).

Notice that because of the numerical factors in front of the $\epsilon$s, the
matrices for $H_1$ and $H_2$ are not equal in general. That means that if we
diagonalize the matrix of $H_1$, the matrix of $H_2$
will not be diagonalized and
will keep the same general form as above. In particular, we can always choose
$H_1$ to be the only doublet that acquires a vev (rotating the doublets will
not change the above form of matrices). Therefore we see that in the quark mass
eigenstate basis, the new Higgs $H_2$ couplings are not diagonal: we have
flavor changing couplings. The nice thing now is that since the flavor
changing couplings are small the stringent experimental limits on flavor
changing neutral currents (FCNC)
actually translate only into lower limits on the mass of new scalars, about
1 TeV, as discussed above.

To avoid problems with large flavor-changing neutral currents, Glashow and
Weinberg \cite{glas77} argued that only one Higgs doublet could couple to
up-type quarks and only one Higgs to down-type quarks. However, this naturality
constraint, known as the Glashow-Weinberg criterion (or natural flavor
conservation (NFC)), was based on an unusual
definition of what is ``natural." For them the avoidance of flavor-changing
neutral currents was natural in a model only if it occured for all values of
the coupling constants of that model.
For us a model will be natural provided
the smallness of any coupling is guaranteed by approximate symmetries
\cite{thoo80}, and we find
that this implies the Glashow-Weinberg criterion is
not necessary (however, see caveat below).

One potential problem arises from the smallness of the observed CP violation,
as noted by Hall and Weinberg\cite{hall93}. The CP violating parameter
$\epsilon_{CP} = Im(\Delta M_K) / \sqrt{2} |\Delta M_K|$ would naively be
expected to be of order unity in our case (we have no reason to assume
{\it a priori} that the Yukawa couplings a real),
contrary to the observed value of
$10^{-3}$. To avoid this problem one might go back to NFC, or, more in the
philosophy of naturalness of small couplings, just say that CP is another
approximately conserved quantity broken by $\epsilon_{CP}$.

\vspace{0.5in}

Here we would like to mention two different limits of the
Yukawa matrices which obey approximate flavor symmetries
(as in (\ref{epsi})), one of which gives NFC, and the other
one which gives democratic matrices.

Notice that if the matrix for $H_2$ is nearly equal to the matrix for $H_1$,
then diagonalization of the $H_1$ matrix will almost diagonalize the second
matrix. In this case the flavor changing couplings become even smaller. This is
of course no surprise, because if the two matrices were exactly equal then only
one linear combination of Higgses ($H_1+H_2$) couples to the quarks:
we have NFC!
This is actually the starting point of Leurer, Nir and Seiberg
\cite{leur93}; i.e., use broken flavor symmetries in combination with weakly
broken NFC.

Also notice that if the Yukawa matrix elements are {\it exactly} \,
a product of $\epsilon$s by which the symmetries are broken, then
the matrices have one large eigenvalue and two eigenvalues equal to zero:
we have flavor democracy(\cite{hara77},\cite{frit94})!
This is easily understood, since this limit
means that only one linear combination of the left handed fields
couples to one linear combination of right handed fields (while the other
two combinations remain massles).

None of these limits follow from the idea of approximate flavor symmetries
alone. Unless they are motivated by a specific model, we must
stay with the general relation of type (\ref{epsi}).

\section{Lepton sector}

By adding the right-handed neutrinos $N_i$, $i=1,2,3$, to the particle content
of the Standard Model we can allow for Dirac type masses. Under the action of
approximate flavor symmetries, whenever an $N_i$ enters a Yukawa interaction,
the corresponding coupling must contain the symmetry breaking parameter
$\epsilon_{N_i}$.

A natural way to justify the smallness of neutrino masses is to use the
{\it see-saw mechanism}\cite{gell79},
in which the smallness of the left-handed neutrino
masses is explained by the new scale of heavy right-handed neutrinos. The mass
matrices will have the structure\cite{rasi94}
\begin{equation}
m_{N_{Dij}} \approx \epsilon_{L_i} \epsilon_{N_j} v_{SM} \, ,
\end{equation}
\begin{equation}
m_{N_{Mij}} \approx \epsilon_{N_i} \epsilon_{N_j} v_{Big} \, ,
\end{equation}
\begin{equation}
m_{E_{ij}} \approx \epsilon_{L_i} \epsilon_{E_j} v_{SM} \, ,
\end{equation}
where $m_{N_D}$ and $m_E$ are the neutrino and charged
lepton Dirac mass matrices,
$m_{N_M}$ is the right-handed neutrino Majorana mass matrix,
$v_{SM}=174 \, {\rm Ge\!V}$
and $v_{Big}$ is the new large mass scale.
The generation indices $i$ and $j$ run from 1 to 3.
In the following we assume a hierarchy in the $\epsilon$s
(i.e. $\epsilon_{L_1} << \epsilon_{L_2} << \epsilon_{L_3}$, etc.)
as suggested by the hierarchy of quark and charged lepton masses.
Then the diagonalization of the neutrino mass matrix will give
a heavy sector with masses
$m_{N_{Hi}} \approx \epsilon^2_{N_i} v_{Big}$ and a very light
sector with mass matrix
\begin{equation}
m_{N_{Lij}} \approx ( m_{N_D} m^{-1}_{N_M} m^T_{N_D} )_{ij}
\approx \epsilon_{L_i} \epsilon_{L_j} \frac {v^2_{SM}} {v_{Big}} \, ,
\end{equation}
where the number
$Tr(\epsilon_N (\epsilon_N \epsilon_N)^{-1} \epsilon_N)$ is assumed to be
of order unity. We have the expected result: the heavy right-handed
neutrino decouples from the theory leaving behind a very
light left-handed neutrino. The masses and mixing angles are
independent of the right-handed symmetry breaking parameters
$\epsilon_{N_i}$:
\begin{eqnarray}
m^N_i
& \approx &
\epsilon^2_{L_i} \frac {v^2_{SM}} {v_{Big}}\  ,
\nonumber\\
m^E_i
& \approx &
\epsilon_{L_i} \epsilon_{E_i} v_{SM}
\ \ \ (no \  sum \  on \  i)\  ,
\nonumber\\
V_{ij}
& \approx &
\frac {\epsilon_{L_i}} {\epsilon_{L_j}}\ \ \  (i<j)\ .
\label{eq:majorana}
\end{eqnarray}
Therefore, besides the unknown scale $v_{Big}$, only two sets of
$\epsilon$s are needed: $\epsilon_{L_i}$ and $\epsilon_{E_i}$.
In fact, the neutrino masses and mixings depend only on $\epsilon_{L_i}$
and they are approximately related through
\begin{equation}
V_{ij} \approx \sqrt{ \frac {m^N_i} {m^N_j} } \, .
\label{eq:relation}
\end{equation}
%
Equation (\ref{eq:relation}) reduces the number of parameters
needed to describe neutrino
masses and mixings by three; for example, given two mixing angles and
one neutrino mass, we can predict the third mixing angle and the other
two neutrino masses. These results are extremely general. They follow simply
from the approximate factorization of the Dirac masses, regardless of
the specific form
of $m^{-1}_{N_M}$, which only contributes to set the scale.

To get further relations one needs some additional information about
the $\epsilon_L$s and $\epsilon_E$s. We tried several plausible
ans\"{a}tze\cite{rasi94} and found that in all
of them the solar neutrino problem
(SNP) can easily be accomodated with
MSW\cite{mikh80} $\nu_e - \nu_{\mu}$ mixing
solution.
However, if we now fix the mass scale from the SNP solution
(requiring $m_{\nu_\mu}$ to be about $10^{-3} {\rm e\!V}$),
then all neutrino masses are too small to close the
Universe. This comes about because the approximate flavor
symmetries tell us that the ratios of neutrino masses are likely to be
of the order of ratios of charged lepton masses (and therefore,
the heaviest neutrino, $\nu_\tau$, is not likely to be heavier
than $1{\rm e\!V}$), as opposed to some proposed quadratic relations.
\section {Acknowledgement}
This work was
supported in part by the Director, Office of
Energy Research, Office of High Energy and Nuclear Physics, Division of
High Energy Physics of the U.S. Department of Energy under Contract
DE-AC03-76SF00098.

\end{document}